\documentclass[aps,onecolumn,prl,groupedaddress,superscriptaddress,amsmath,amssymb,amsthm]{revtex4}
\usepackage{subfigure,hyperref,bbm,times}
\usepackage[T1]{fontenc}
\usepackage{braket}
\usepackage{amsthm}
\usepackage{epsfig}
\usepackage{color}
\usepackage{graphicx}
\usepackage{dcolumn}
\usepackage{bm}

\providecommand{\openone}{\leavevmode\hbox{\small1\kern-3.8pt\normalsize1}}

\usepackage{soul}

\begin{document}
\title{Effects of indistinguishability in a system of three identical qubits}

\author{Alessia Castellini}
\email{alessia.castellini@unipa.it}
\affiliation{Dipartimento di Fisica e Chimica, Universit\`a di Palermo, via Archirafi 36, 90123 Palermo,
Italy}

\author{Rosario Lo Franco}
\email{rosario.lofranco@unipa.it}
\affiliation{Dipartimento di Fisica e Chimica, Universit\`a di Palermo, via Archirafi 36, 90123 Palermo, Italy}
\affiliation{Dipartimento di Energia, Ingegneria dell'Informazione e Modelli Matematici, Universit\`{a} di Palermo, Viale delle Scienze, Edificio 9, 90128 Palermo, Italy}

\author{Giuseppe Compagno}
\email{giuseppe.compagno@unipa.it}
\affiliation{Dipartimento di Fisica e Chimica, Universit\`a di Palermo, via Archirafi 36, 90123 Palermo, Italy}

\date{\today }

\begin{abstract}
Quantum correlations of identical particles are important for quantum-enhanced technologies. The recently introduced non-standard approach to treat identical particles (\textit{G. Compagno et al., Phil. Trans. R. Soc. A 376, 20170317 (2018)}) is here exploited to show the effect of particle indistinguishability on the characterization of entanglement of three identical qubits. 
We show that, by spatially localized measurements in separated regions, three independently-prepared separated qubits in a pure elementary state behave as distinguishable ones, as expected. On the other hand, delocalized measurements make it emerge a measurement-induced entanglement. We then find that three independently-prepared boson qubits under complete spatial overlap exhibit genuine three-partite entanglement.  These results evidence the effect of spatial overlap on identical particle entanglement and show that the latter depends on both the quantum state and the type of measurement.

\end{abstract}



\maketitle

\section{Introduction}
Identical particles (e.g., photons, atoms, electrons, qubits) are the building blocks of quantum networks and quantum-enhanced devices \cite{Ladd(2010),Braun(2018)}. As a result, understanding and exploiting their physical properties and their quantum correlations have both a conceptual and practical relevance.
In the standard quantum mechanical approach, identical particles are labeled with unphysical labels and treated like this from the beginning as if they were not identical. Their quantum states are then  (anti)symmetrized with respect to labels that gives them the structure of entangled states. 

Attempts have been done to characterize this entanglement coming from label (anti)symmetrization. 
Such an entanglement, when identical particles are independently prepared and spatially separated, is believed to be not usable being only the result of the standard formalism which employs unphysical labels \cite{Peres(1995)}. Even in the presence of spatial overlap, this entanglement is considered purely formal \cite{Ghirardi(2004),Tichy(2011)}. Moreover, within the standard formalism, it is also impossible to adopt the common tools, such as partial traces and von Neumann entropy, to quantify the amount of entanglement, since identical particles are individually unaddressable \cite{Ghirardi(2004)}.

A particle-based non-standard approach (NSA) has been introduced \cite{LoFranco(2016)}, which does not use unobservable labels to describe identical particle quantum states. It avoids from the very beginning the problem of entanglement arising from unphysical labels and paves the way to an advantageous study of quantum correlations in systems of identical particles. NSA has been originally presented for two-particle systems \cite{LoFranco(2016)} and has allowed to prove that the Schmidt decomposition is also applicable to identical particles \cite{Sciara(2017)}. The NSA formalism has been successively generalized to the case of an arbitrary number of identical particles and its connection with second quantization has also been shown \cite{Compagno(2018)}. In addition, NSA allows us to use the tools commonly used to quantify entanglement for distinguishable particles. More recently, this approach has led to the definition of a new operational framework in order to exploit, for quantum teleportation, the entanglement due to the indistinguishability of two independently prepared qubits \cite{LoFranco(2018)}. 

We here apply the generalized NSA formalism to specifically treat entanglement in systems with more than two identical particles. In particular, we show the effect of indistinguishability for states of three identical qubits also when they spatially overlap, comparing it to the case of non-identical qubits.

\begin{figure*}[!t]
\centering
\includegraphics[scale=0.43]{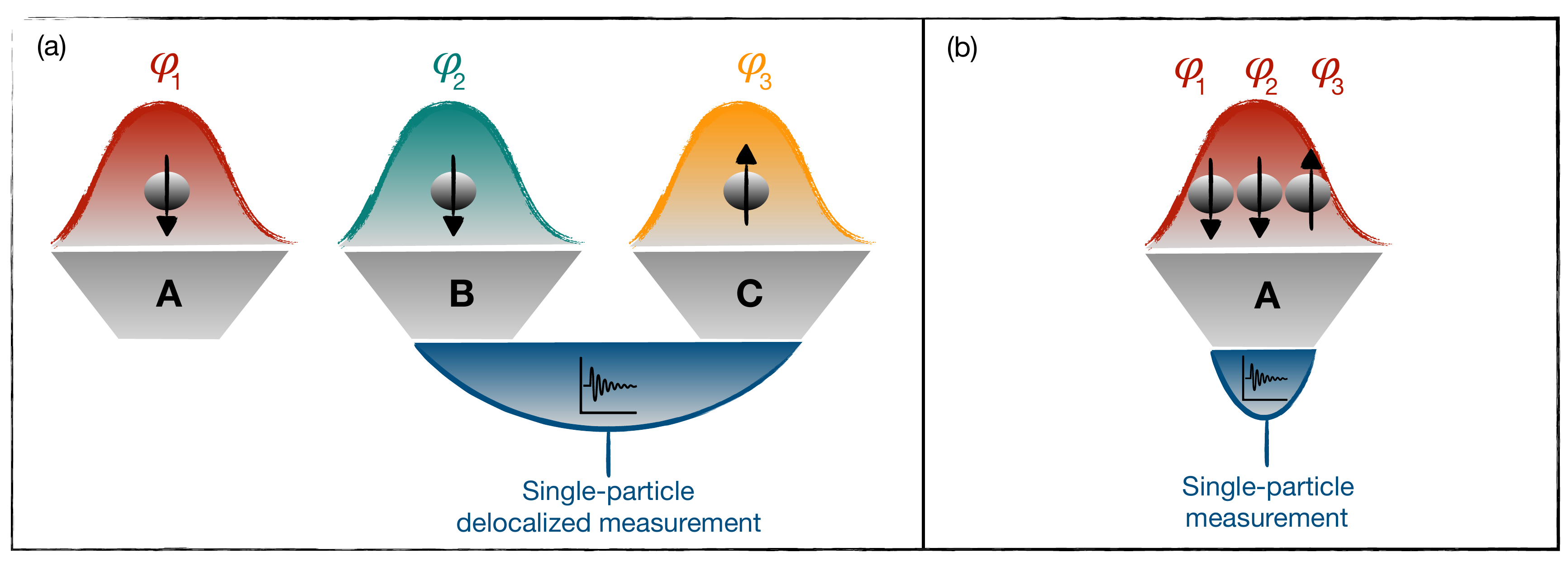}
\caption{(a) Configuration associated to the state $|\Phi^{(3)}\rangle$ when particles are in separate spatial regions (grey zones). (b) Complete spatial overlap of three qubits in the spatial region A.}
\label{Figura}
\end{figure*}

\section{Results}
Let us consider three identical qubits, each in the state $|\varphi_k \ \sigma_k\rangle\equiv|\varphi_k\rangle\otimes|\sigma_k\rangle$ ($k=1,2,3$), where $\varphi_k$ is the $k$-th spatial wavefunction and $\sigma_k$ is the corresponding pseudospin $\uparrow$, $\downarrow$ (e.g., components of a spin-$1/2$ fermion, two energy levels of a boson, horizontal $H$ and vertical $V$ polarizations of a photon). Using the NSA \cite{Compagno(2018)}, the three-qubit elementary state is $|\Phi^{(3)}\rangle=|\varphi_1 \ \sigma_1,\varphi_2 \ \sigma_2,\varphi_3 \ \sigma_3\rangle$, which cannot be separated in terms of tensor products of single-qubit states: it must be meant as an overall object representing a list of the single-particle states. 

We start taking the total system in the pure state 
$|\Phi^{(3)}\rangle=|\mathrm{A\downarrow,B\downarrow,C\uparrow}\rangle$, where the three spatial wavefunctions have support in the three corresponding separated spatial regions $\mathrm{A}$, $\mathrm{B}$ and $\mathrm{C}$, as depicted in Fig. \ref{Figura}(a). Considering all the possible bipartitions $\mathrm{(AB)}$-$\mathrm{C}$, $\mathrm{(CA)}$-$\mathrm{B}$, $\mathrm{(BC)}$-$\mathrm{A}$ of the three-particle system, by performing localized projective measurements and partial traces \cite{Compagno(2018)} one finds that all the three bipartitions give pure one-particle and two-particle reduced density matrices. Following conventional wisdom this state is, as expected, fully uncorrelated \cite{horodeckireview}. Thus, separated identical particles, addressed by spatially localized measurements, behave analogously to distinguishable particles. However, in contrast with the case of distinguishable particles which are individually addressable, quantum correlations between identical particles depend on the region where measurements are performed. In fact, if partial traces of an identical particle state are made onto non-local basis, a ``measurement-induced'' entanglement \cite{LoFranco(2016)} is obtained. For example, by making on our system of Fig. \ref{Figura}(a) in the state $\rho^{(3)}=|\Phi^{(3)}\rangle \langle \Phi^{(3)}|$ a non-local one-particle trace onto the delocalized basis $\Bigl\{\dfrac{1}{\sqrt{2}}|(\mathrm{B+C})\downarrow\rangle, \dfrac{1}{\sqrt{2}}|(\mathrm{B+C})\uparrow\rangle\Bigr\}$, we obtain the two-particle mixed state $\rho^{(2)}=\dfrac{1}{2}(\mathrm{|A\downarrow,C\uparrow\rangle\langle A\downarrow,C\uparrow|+|A\downarrow,B\downarrow\rangle\langle A\downarrow,B\downarrow|})$. This shows how entanglement of spatially separated identical particles is affected by the non-local nature of measurements.

We now compare the above results with the analogous case of distinguishable qubits. The global state can be written as the tensor product of single-particle labeled states, namely $|\Phi^{(3)}_\mathrm{d}\rangle=\mathrm{|A\downarrow\rangle_1\otimes|B\downarrow\rangle_2 \otimes |C\uparrow\rangle_3}$ (where labels 1, 2, and 3 identify distinguishable particles). This state is manifestly separable and so unentangled \cite{horodeckireview}.
Moreover, if we make partial traces over a non-local basis (e.g., $\mathcal{B}_k=\Bigl\{\dfrac{1}{\sqrt{2}}|(\mathrm{A+B})\downarrow\rangle, \dfrac{1}{\sqrt{2}}|(\mathrm{A+B})\uparrow\rangle\Bigr\}_k$), due to the fact that now measurements are made on an individual particle, pure reduced density matrices are obtained: the particles thus remain fully uncorrelated. This result does not change even when the three distinguishable qubits completely spatially overlap (A$=$B$=$C). This aspect highlights the difference on the entanglement properties of independently-prepared identical and non-identical particles.  

To have entanglement of three spatially separated qubits by local measurements, linear combinations of elementary states \cite{Compagno(2018)} are necessary. In particular, we take a superposition of two three-qubit states corresponding to the well-known GHZ state of distinguishable particles \cite{horodeckireview,Amico(2008)}, $|\Lambda^{(3)}_\mathrm{GHZ}\rangle=\frac{1}{\sqrt{2}}(|\mathrm{A}\downarrow,\mathrm{B}\downarrow,\mathrm{C}\downarrow\rangle+|\mathrm{A}\uparrow,\mathrm{B}\uparrow,\mathrm{C}\uparrow\rangle)$. All the distinct bipartitions, by means of local partial traces, are easily seen to produce totally mixed reduced density matrices: therefore, the state is maximally three-partite entangled. A further advantage of the NSA formalism here emerges because, when the $|\Lambda^{(3)}_\mathrm{GHZ}\rangle$ is expressed in terms of unphysical labels, it consists of a linear combination of twelve different product states of three one-particle state vectors. This would make the analysis technically  cumbersome and would hinder the identification of physical entanglement from the unphysical one associated to labels.

To evidence the role of the spatial overlap, we now consider the state of three boson qubits condensed in the same spatial mode $\mathrm{A}$, as illustrated in Fig. \ref{Figura}(b).
This state, using the NSA formalism \cite{Compagno(2018)}, has the form $|\Psi^{(3)}\rangle=\frac{1}{\sqrt{2}}|\mathrm{A\downarrow,A\downarrow,A\uparrow}\rangle$ and admits only one bipartition (two particles)-(one particle). By performing partial traces onto this bipartition, one obtains the one-particle and two-particle reduced density matrices 
\begin{equation}
\rho^{(1)}=\dfrac{1}{\tilde{\mathcal{N}}}\mathrm{Tr}^{(2)}|\Psi^{(3)}\rangle\langle\Psi^{(3)}|=\frac{2}{3}|\mathrm{A}\downarrow\rangle\langle \mathrm{A}\downarrow|+\frac{1}{3}|\mathrm{A}\uparrow\rangle\langle \mathrm{A}\uparrow|,
\end{equation}
\begin{equation}
\rho^{(2)}=\dfrac{1}{\mathcal{N}}\mathrm{Tr}^{(1)}|\Psi^{(3)}\rangle\langle\Psi^{(3)}|=\frac{2}{3}|\mathrm{A}\uparrow,\mathrm{A}\downarrow\rangle\langle \mathrm{A}\uparrow,\mathrm{A}\downarrow|+\frac{1}{3}\frac{|\mathrm{A}\uparrow,\mathrm{A}\uparrow\rangle}{\sqrt{2}}\frac{\langle \mathrm{A}\uparrow,\mathrm{A}\uparrow|}{\sqrt{2}}.
\end{equation}
These density matrices are manifestly mixed and give the same von Neumann entropy  $S(\rho^{(1)})=S(\rho^{(2)})=-\dfrac{1}{3}\mathrm{log}_2 \dfrac{1}{3}-\dfrac{2}{3}\mathrm{log}_2 \dfrac{2}{3}$. This result, differently from the uncorrelated state $|\Phi^{(3)}\rangle$, witnesses that $|\Psi^{(3)}\rangle$ is three-partite entangled \cite{horodeckireview}.

\section{Discussion}
We have analyzed the role of spatial overlap, and thus of particle indistinguishability, on the entanglement of three identical qubits. We have found that, by making spatially localized measurements in separated regions, three independently-prepared separated qubits in a pure elementary state behave as distinguishable ones, as expected \cite{facchi}. Spatially separated identical particles are entangled only if they are prepared in a linear combination of elementary states. Instead, if a delocalized basis is chosen to make partial traces, mixed reduced density matrices are obtained. This confirms that quantum correlations of identical particles are not an intrinsic property of the quantum state alone but also depend on the spatial region where measurements are performed \cite{LoFranco(2016),LoFranco(2018)}. 

We have then considered three bosons qubits in the configuration of complete spatial overlap. In this case, spatial overlap constitutes an entangling mechanism \cite{LoFranco(2018)}: two mixed reduced density matrices are associated to the sole possible bipartition (two-particle)-(one particle) and testify a genuine three-partite entanglement. 

The non-standard approach to identical particles \cite{LoFranco(2016),Sciara(2017),Compagno(2018)} allows a clear identification of the physical role of particle indistinguishability in the characterization of quantum correlations. This opens the way to a deeper understanding of the collective properties of identical particle systems and to the operational use of indistinguishability in other contexts such as resource theory of coherence \cite{Winter(2016)} and quantum metrology \cite{Giovannetti(2011)}.

\end{document}